\documentclass[runningheads]{llncs}
\usepackage[T1]{fontenc}
\usepackage{graphicx}
\usepackage{amsmath}
\usepackage{amssymb}
\usepackage{url}
\usepackage{array}
\usepackage{booktabs}
\usepackage{multirow}
\begin{document}
\title{On-Policy Self-Distillation for Multi-Dialect ASR: Mastering Dialects, Retaining Mandarin}
\titlerunning{OPSD for Multi-Dialect ASR}
%
\author{Shuiyuan Wang\inst{1} \and
Bingshen Mu\inst{1} \and
Pengshen Zhang\inst{2} \and
Chengyou Wang\inst{1} \and
Yujie Liao\inst{1} \and
Chengdong Liang\inst{2} \and
Binbin Zhang\inst{2} \and
Qiangze Feng\inst{3} \and
Lei Xie\inst{1}}
\authorrunning{S. Wang et al.}
\institute{Audio, Speech and Language Processing Group (ASLP@NPU), School of Computer Science, Northwestern Polytechnical University, Xi’an, China \and
WeNet Community \and
NEXDATA TECHNOLOGY INC.\\
\email{wangshuiyuan@mail.nwpu.edu.cn, lxie@nwpu.edu.cn}}
%
%
\maketitle
\begin{abstract}
Recent large-scale ASR models already achieve strong Mandarin recognition accuracy and have some ability to recognize Chinese dialects. However, their dialect recognition accuracy is still limited in real-world speech. Direct dialect adaptation can lower dialect CER, but it may also raise Mandarin CER. We therefore study how to adapt a capable ASR model to improve multi-dialect recognition without degrading Mandarin recognition. We adopt an adaptation pipeline where continual pre-training (CPT) and dialect supervised fine-tuning (SFT) provide a strong foundation, and On-Policy Self-Distillation (OPSD) serves as the final refinement. OPSD addresses the train--test mismatch in autoregressive ASR by training the student model on its own decoded prefixes while a frozen teacher, conditioned on the reference transcript as privileged context, provides soft token-level targets. This replaces hard cross-entropy updates on dialect data with distillation, preserving Mandarin ability while refining dialect recognition. We instantiate the framework with Qwen3-ASR-1.7B and evaluate it on public and internal Mandarin and dialect test sets. Under matched refinement data and schedule, OPSD improves dialect recognition without raising Mandarin CER, whereas continued teacher-forced fine-tuning increases Mandarin CER. We will release the model weights and evaluation scripts.\footnote{\url{https://github.com/ASLP-lab/CN-MultiDialect-ASR}}

\end{abstract}
\section{Introduction}

Recent large-scale ASR models~\cite{radford2023whisper,zhang2023usm,pratap2024mms,chu2023qwen,bai2024seed,shang2025firered,shi2026qwen3asr} offer a capable foundation for Mandarin and dialect ASR~\cite{prabhavalkar2023survey}, with strong Mandarin performance and some ability to recognize dialects. The remaining challenge is to improve their dialect robustness without sacrificing the Mandarin ability they already provide.

However, dialect speech is still difficult to recognize in real-world applications~\cite{chen2025polyspeech}. Different regions use different dialects, and many dialects have little labeled training data. As a result, current ASR models still perform poorly on dialect speech in real-world scenarios.

The adaptation objective therefore has two parts. Fine-tuning with more dialect speech can lower dialect character error rate (CER), but it may also increase Mandarin CER. The model must gain robustness to dialect variation while keeping its strong Mandarin recognition ability. This trade-off motivates the staged adaptation framework described below.

We adopt a three-stage adaptation framework for this trade-off. First, continual pre-training (CPT) further adapts the strong base model to large-scale Mandarin-dialect data. Second, dialect supervised fine-tuning (SFT) increases the training weight of dialect speech while retaining a small amount of Mandarin data. This stage is intended to lower dialect CER, but it can still increase Mandarin CER. Third, we apply On-Policy Self-Distillation (OPSD)~\cite{opsd2026} as the final refinement objective to lower dialect CER under decoding states closer to inference while preserving Mandarin performance.

OPSD is well suited for autoregressive ASR refinement because it replaces teacher-forced prefixes with student-decoded ones, reducing the mismatch between training and inference. In teacher-forced training, the model predicts the next token from gold transcript prefixes. At inference time, it must predict from its own previous outputs. OPSD closes part of this gap by using prefixes generated by the student itself. At inference time, only the student is used.

We instantiate this framework with Qwen3-ASR-1.7B~\cite{shi2026qwen3asr}, an open-source model that already has strong Mandarin recognition and some dialect capability. The experiments compare the effects of the three adaptation stages. Direct dialect SFT lowers dialect CER but substantially increases Mandarin CER, showing the trade-off caused by dialect-focused training. CPT followed by dialect SFT improves dialect performance with a smaller Mandarin cost. Starting from this checkpoint, OPSD lowers CER on both Mandarin and dialect test sets. Compared with continuing teacher-forced fine-tuning on the same refinement data, OPSD achieves broader improvements across the evaluation sets.

Our contributions are summarized as follows:
\begin{itemize}
    \item We adopt a three-stage framework for Chinese multi-dialect ASR adaptation, consisting of CPT, dialect SFT, and OPSD refinement.
    \item We propose OPSD for autoregressive ASR refinement, where the reference transcript serves as training-time privileged teacher context and student-decoded prefixes provide on-policy states.
    \item We provide a controlled empirical study on public and internal Mandarin and dialect test sets. The proposed framework improves dialect recognition while maintaining or improving Mandarin recognition, and we will release the model weights and evaluation scripts to support future research.
\end{itemize}

\section{Related Work}

\subsection{Chinese Dialect Speech Data}
Recent public corpora have expanded the coverage of Chinese dialect speech. WenetSpeech-Yue~\cite{li2025wenetspeechyue}, WenetSpeech-Chuan~\cite{dai2025wenetspeechchuan}, and WenetSpeech-Wu~\cite{shi2025wenetspeech} provide large-scale transcribed speech for Cantonese, Sichuan, and Wu, respectively. KeSpeech~\cite{tang2022kespeech} covers Mandarin together with several Chinese subdialects. These resources support training and evaluation for several major dialect families.

Although more dialect data are now available, the coverage is still uneven. Some dialects have large transcribed corpora, while many local varieties have limited labeled data. Differences in speakers, recording conditions, and speech domains also make dialect speech difficult to recognize in real-world applications~\cite{hinsvark2024domain,he2024star}. Therefore, existing resources support research on several major dialects, but they do not fully represent the range of Chinese dialect speech.

\subsection{Chinese Dialect ASR Models}
Several modeling approaches have been proposed for Chinese dialect ASR. Dolphin-CN-Dialect~\cite{meng2026dolphin} is a Chinese dialect ASR model that focuses on dialect-rich training and evaluation. Mixture-of-experts models use shared components and different experts to model multiple dialects~\cite{li2022multitask}. Adapter methods add small trainable modules to a pretrained ASR model, which makes adaptation more efficient for long-tail languages or dialects~\cite{bai2024adapter}. Other studies use dialect-related representations or embeddings to provide the model with dialect information~\cite{chang2026dialectemb}.

These methods improve dialect modeling through model structure, small adaptation modules, or dialect information. However, dialect adaptation may also affect the model's Mandarin ability. A model for Chinese multi-dialect ASR must therefore learn dialect variation while keeping its existing Mandarin performance. This issue is closely related to the choice of adaptation objective.

\subsection{Adaptation Objectives for ASR}
Adapting a pretrained ASR model usually starts with supervised fine-tuning on target speech. This objective is simple and effective. However, it also moves the model toward the target distribution. In Chinese multi-dialect ASR, the goal is not only lower dialect CER. The model should also keep the Mandarin ability learned from much larger data. This makes dialect adaptation close to continual learning~\cite{li2021lwf,wang2024ewc}: the model learns new varieties while an existing ability must be retained.

Several lines of work address parts of this issue. Parameter-efficient methods, such as adapters and LoRA~\cite{hu2022lora}, reduce the cost of adaptation. Continual-learning methods add constraints so the model does not move too far from earlier tasks~\cite{li2021lwf,wang2024ewc}. Domain adaptation and self-training use unlabeled or weakly labeled speech to improve robustness~\cite{ahmad2024progressive}. These studies show that adaptation is not only a data problem. The training objective also affects which ability is strengthened and which ability is weakened.

Distillation provides another way to guide the adapted model. In ASR, teacher--student training can use stronger models, multiple teachers, pseudo labels, or consistency targets~\cite{yang2023multifoundation,higuchi2021momentum,zhang2024crctc}. Soft targets retain more of a capable teacher's knowledge than one-hot labels, which can help limit overwrite of an existing ability such as Mandarin recognition. Soft targets alone are not enough, however. Standard supervised fine-tuning still trains the decoder on reference transcript prefixes, while autoregressive decoding must continue from the model's own previous tokens. This train--test gap has long been studied for sequence models, for example through scheduled sampling that mixes gold and model prefixes during training~\cite{bengio2015scheduled}. Recent on-policy distillation further trains the student on its own rollouts while a teacher provides dense token-level feedback on those states~\cite{agarwal2024gkd}.

This train--test gap is especially harmful for dialect speech: early errors are more common, and a wrong prefix quickly leaves the gold path. Continued hard cross-entropy on dialect data can further push the model away from Mandarin. An effective final objective should therefore train under decoding states closer to inference, while using soft teacher guidance to refine difficult dialect cases without another aggressive one-hot shift.

\section{Method}

\subsection{Overview}

\begin{figure}[t]
\centering
\includegraphics[width=\columnwidth]{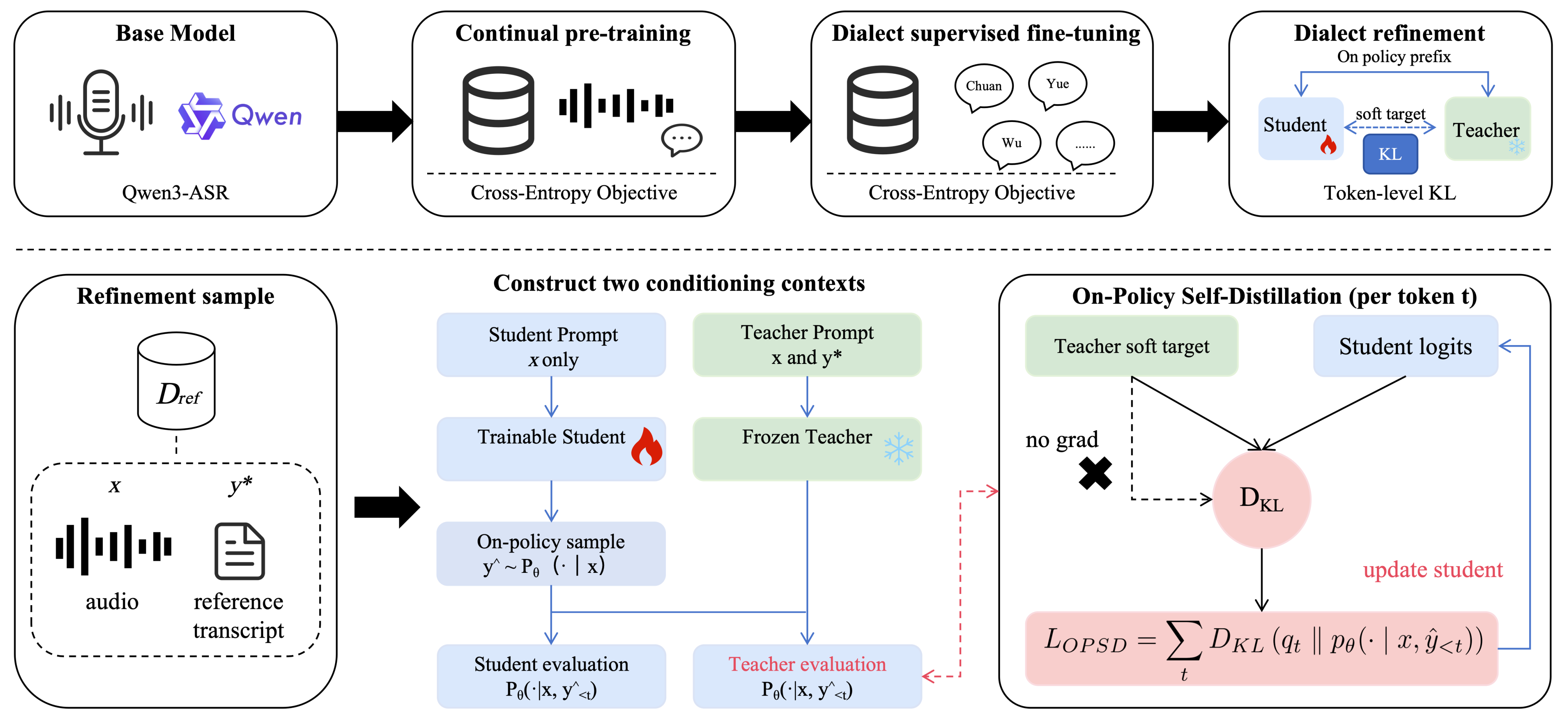}
\caption{Overview of the staged adaptation pipeline. Top: from the base model through CPT and SFT to OPSD. Bottom: OPSD with student on-policy prefixes, a frozen teacher with reference transcript as privileged context, soft targets~$q_t$, and token-level KL.}
\label{fig:framework}
\end{figure}

Figure~\ref{fig:framework} shows the proposed framework. The pipeline has three stages. CPT builds a stronger ASR foundation. SFT shifts the training mixture toward dialect speech and lowers dialect CER, but may increase Mandarin CER. OPSD then refines the SFT checkpoint to reduce this Mandarin CER increase.

\begin{enumerate}
    \item \textbf{CPT:} continual pre-training on large-scale Mandarin-dialect data. This step strengthens the overall ASR foundation.
    \item \textbf{SFT:} Dialect supervised fine-tuning. This step lowers dialect CER, but it may also increase Mandarin CER.
    \item \textbf{OPSD:} Final refinement on selected dialect training data. This step further trains the SFT checkpoint with the OPSD objective.
\end{enumerate}

\subsection{CPT Stage}

The first stage is CPT. It adapts the base ASR model to large-scale Mandarin-dialect data before the dialect SFT stage. The model is trained on a large Mandarin-dialect corpus
\[
\mathcal{D}_{\text{zh}}=\{(\mathbf{x}_i,\mathbf{y}_i)\}_{i=1}^{N_{\text{zh}}},
\]
This dataset comprises a union of large-scale Mandarin-dialect corpora. Detailed sources and hour counts are given in Table~\ref{tab:datasets}. The loss is standard sequence-level cross-entropy:

\begin{equation}
\mathcal{L}_{\text{CPT}} = -\sum_{i=1}^{N_{\text{zh}}} \sum_{t=1}^{T_i} \log p_\theta(y_{i,t} \mid \mathbf{x}_i, \mathbf{y}_{i,<t}),
\label{eq:stage1}
\end{equation}

where $p_\theta$ denotes the ASR model parameterized by $\theta$. This stage uses paired audio and transcripts. It provides a stronger ASR foundation for the subsequent dialect SFT stage.

\subsection{SFT Stage}
The second stage is dialect SFT. It lowers dialect CER on the CPT checkpoint. This stage does not mainly add new dialect types. Instead, it changes the sampling ratio and gives dialect speech a larger weight. We construct
\[
\mathcal{D}_{\text{zh}}^{\text{dial}}=\mathcal{D}_{\text{dial}}^{\text{all}} \cup \mathcal{D}_{\text{mand}}^{\text{small}},
\]
where $\mathcal{D}_{\text{dial}}^{\text{all}}$ contains all dialect training data already used in CPT. $\mathcal{D}_{\text{mand}}^{\text{small}}$ is a small amount of Mandarin training data. The stage uses the same sequence-level cross-entropy objective as CPT.
Compared with $\mathcal{D}_{\text{zh}}$, the main change is the Mandarin--dialect sampling ratio. Corpus sources and hour counts are given in Section~4.1. This stage lowers dialect CER. It also produces the shared starting checkpoint for the final refinement stage.

\subsection{OPSD Stage}
The third stage applies OPSD to the SFT checkpoint. The goal is twofold. First, we train under decoding states closer to inference, which helps dialect speech: standard teacher-forced SFT predicts each token from the reference transcript prefix, but decoding must continue from the model's own previous tokens, and an early error can leave the gold path. Second, we replace another hard cross-entropy update on dialect data with soft targets from a capable teacher, which helps limit further overwrite of Mandarin ability.

We formulate OPSD for autoregressive ASR refinement. In the original formulation, a single model acts as both teacher and student with different contexts, and gradients do not flow through the teacher pathway. For ASR refinement, we instead keep a frozen teacher copy $\bar{\theta}$ initialized from the SFT checkpoint, and train only the student $\theta$. This keeps the soft targets stable while the student updates on long speech sequences. The refinement set $\mathcal{D}_{\text{ref}}$ is selected from $\mathcal{D}_{\text{dial}}^{\text{all}}$.

For each $(\mathbf{x}, \mathbf{y}^\star)\in\mathcal{D}_{\text{ref}}$, the student samples its own hypothesis, and the teacher provides token-level soft targets on the same student prefixes. The teacher receives the reference transcript as privileged context; the student does not. We write this privileged context as $c(\mathbf{y}^\star)$. It is a training-time prompt field separated from the student prefix by a delimiter. It is used only in the teacher pathway and is unavailable to the student or the deployed model. For each training example:
\begin{enumerate}
    \item Sample a student hypothesis $\hat{\mathbf{y}}_{\leq T}$ from $p_\theta(\cdot \mid \mathbf{x})$ with the standard ASR prompt. Here $T$ is the number of sampled prediction positions up to the EOS token. We use temperature $\tau=0.8$ during training.
    \item For each position $t=1,\ldots,T$, compute the teacher distribution on the same student prefix:
\begin{equation}
q_t(v) = p_{\bar{\theta}}(v \mid \mathbf{x}, c(\mathbf{y}^\star), \hat{\mathbf{y}}_{<t}), \quad v \in \mathcal{V}.
\label{eq:teacher}
\end{equation}
    \item Update only the student by matching this distribution:
\begin{equation}
\mathcal{L}_{\text{OPSD}} = \sum_{t=1}^{T} D_{\text{KL}}\big(q_t \,\|\, p_\theta(\cdot \mid \mathbf{x}, \hat{\mathbf{y}}_{<t})\big).
\label{eq:kl}
\end{equation}
\end{enumerate}
Only $\theta$ is updated. The teacher $\bar{\theta}$ remains frozen. At inference, $\mathbf{y}^\star$ and $c(\mathbf{y}^\star)$ are absent. The deployed model is the student pathway $p_\theta(\cdot \mid \mathbf{x}, \hat{\mathbf{y}}_{<t})$. We use pure $\mathcal{L}_{\text{OPSD}}$ without an auxiliary cross-entropy term.

\section{Experimental Setup}

\subsection{Datasets}

The full training collection $\mathcal{D}_{\text{zh}}$ contains approximately 100k hours of Mandarin-dialect data. Table~\ref{tab:datasets} lists each source and its approximate hours after filtering. CPT trains on this full collection. Dialect SFT keeps the same sources but increases the sampling weight of dialect speech and retains a small Mandarin subset, forming $\mathcal{D}_{\text{zh}}^{\text{dial}}$.

For OPSD and Continued SFT, we further select a refinement set $\mathcal{D}_{\text{ref}}$ of approximately 5k hours from the dialect training partition of $\mathcal{D}_{\text{dial}}^{\text{all}}$. We decode these utterances with the SFT model, compute CER against the references, and filter out cases with unreliable metadata, abnormal duration, empty or extremely short transcripts, or failed audio loading. From the remainder, we keep high-CER utterances from human-annotated sources and cap the hours per dialect so that a few difficult dialects do not dominate. No development or test utterances are used. When speaker IDs are available, test speakers are held out from training and refinement; we also apply utterance-level de-duplication across training, refinement, development, and test sets.

We evaluate on 13 public and shared Mandarin and dialect test sets. Eight are Mandarin: AISHELL-1, AISHELL-2, KeSpeech, SpeechIO-1 to SpeechIO-3~\cite{song2024touchasp}, and the WenetSpeech Test\_Meeting and Test\_Net sets. Five are dialect sets: WenetSpeech-Yue (long/short), WenetSpeech-Chuan (easy/hard), and WenetSpeech-Wu. We further evaluate an Internal Dialect suite; the dialect names appear in Table~\ref{tab:dialect}.

\begin{table}[t]
\caption{Training data sources, dialect coverage, and approximate hours.}
\label{tab:datasets}
\centering
\scriptsize
\setlength{\tabcolsep}{1.5pt}
\begin{tabular}{>{\raggedright\arraybackslash}p{4.5cm}@{\hspace{3pt}}>{\raggedright\arraybackslash}p{6.3cm}@{\hspace{0pt}}>{\raggedleft\arraybackslash}p{1.0cm}}
\toprule
\textbf{Source} & \textbf{Dialect coverage} & \textbf{Hours} \\
\midrule
WenetSpeech~\cite{zhang2022wenetspeech} & Mandarin & $\sim$22.4k \\
AISHELL-1~\cite{bu2017aishell} & Mandarin & $\sim$178 \\
AISHELL-2~\cite{du2018aishell2} & Mandarin & $\sim$1,000 \\
AliMeeting~\cite{yu2021m2met} & Mandarin & $\sim$0.1k \\
Common Voice 17.0~\cite{ardila2020commonvoice} & Mandarin & $\sim$234 \\
MAGICDATA Read Speech~\cite{yang2022magicdata} & Mandarin & $\sim$755 \\
KeSpeech~\cite{tang2022kespeech} & Accented Mandarin & $\sim$1.5k \\
WenetSpeech-Yue~\cite{li2025wenetspeechyue} & Cantonese & $\sim$21.8k \\
WenetSpeech-Chuan~\cite{dai2025wenetspeechchuan} & Sichuan & $\sim$10.0k \\
WenetSpeech-Wu~\cite{shi2025wenetspeech} & Wu & $\sim$8.0k \\
\multirow{5}{4.5cm}{\raggedright Internal data} & Mandarin, Anhui, Cantonese, Changsha, Chaoshan & \multirow{5}{1.0cm}{\raggedleft$\sim$34.1k} \\
& Dongbei, Henan, Kejia, Minnan & \\
& Nanchang, Nanjing, Shanxi, Shaanxi & \\
& Shandong, Shanghai, Sichuan, Suzhou & \\
& Tianjin, Wuhan, Xuzhou, Ningxia, Gansu & \\
\midrule
Total training data & & $\sim$100k \\
\bottomrule
\end{tabular}
\end{table}

\subsection{Baselines}

We compare against three open-source ASR models: GLM-ASR-Nano-2512~\cite{zeng2024glm4voice}, Fun-ASR-Nano-2512~\cite{gao2023funasr}, and Qwen3-ASR-1.7B. Hereafter we refer to them as GLM-ASR, Fun-ASR, and Qwen3-ASR. Qwen3-ASR is also the starting checkpoint of our pipeline. The labels CPT, SFT, and OPSD denote the cumulative checkpoints after each stage.

For controlled ablations in Section~\ref{sec:ablation}, we add two settings:
\begin{itemize}
    \item \textbf{Direct Dialect SFT:} the same $\mathcal{D}_{\text{zh}}^{\text{dial}}$ and training recipe as SFT, but starting from Qwen3-ASR without CPT.
    \item \textbf{Continued SFT:} the SFT checkpoint further trained on $\mathcal{D}_{\text{ref}}$ with teacher-forced cross-entropy, using the same data and schedule as OPSD.
\end{itemize}

\subsection{Evaluation Metrics}

We use Character Error Rate (CER) as the primary evaluation metric for all test sets. All references and hypotheses are scored at the character level after shared text normalization. The pipeline unifies full-width and half-width characters. It removes punctuation and whitespace. It also converts common spoken variants to a canonical written form when the corpus guidelines provide one. Dialect utterances are transcribed in Chinese characters rather than phonetic dialect orthography. Code-switched Mandarin--dialect spans are scored as one character sequence against the reference. We do not apply dialect-specific pronunciation lexicons at scoring time.

Macro-averages treat each test set equally and are not weighted by utterance count. We report four CER averages as defined in Table~\ref{tab:avg_definition}. \textbf{Overall Avg.} is the macro-average of all 31 test sets.

\begin{table}[t]
\caption{Definition of the four CER averages. Each average is a simple macro-average of its constituent test sets.}
\label{tab:avg_definition}
\centering
\scriptsize
\setlength{\tabcolsep}{1pt}
\begin{tabular}{>{\raggedright\arraybackslash}p{2.0cm}>{\raggedright\arraybackslash}p{10.0cm}}
\toprule
\textbf{Average} & \textbf{Constituent test sets} \\
\midrule
Mandarin Avg. & AISHELL-1, AISHELL-2, KeSpeech, SpeechIO-1$\sim$3, Test\_Meeting, Test\_Net \\[3pt]
Dialect Avg. & WenetSpeech-Yue, WenetSpeech-Chuan, WenetSpeech-Wu \\[3pt]
\multirow{3}{2.0cm}{\raggedright Internal Avg.} & Anhui, Cantonese, Changsha, Chaoshan, Dongbei, Henan \\
& Kejia, Minnan, Nanchang, Nanjing, Shanxi, Shaanxi \\
& Shandong, Shanghai, Sichuan, Suzhou, Wuhan, Xuzhou \\[3pt]
Overall Avg. & All of the above \\
\bottomrule
\end{tabular}
\end{table}

\subsection{Implementation Details}
\label{sec:impl}

We use Qwen3-ASR as the base model. All stages are trained on 8 NVIDIA RTX A6000 GPUs with DeepSpeed ZeRO-2 and FlashAttention-2. CPT runs for one epoch on $\mathcal{D}_{\text{zh}}$. It uses learning rate $1\times10^{-5}$, global batch size 1536, and standard cross-entropy loss. SFT starts from the CPT checkpoint and runs for one epoch on $\mathcal{D}_{\text{zh}}^{\text{dial}}$. It uses the same learning rate, batch size, and objective. Direct Dialect SFT uses the same $\mathcal{D}_{\text{zh}}^{\text{dial}}$ and hyperparameters as SFT, but starts from Qwen3-ASR without CPT.

Stage-3 refinement compares OPSD with Continued SFT under a matched schedule. Both start from the same SFT checkpoint. Both train for one epoch on the same $\sim$5k-hour dialect data $\mathcal{D}_{\text{ref}}$. The learning rate is $1\times10^{-4}$ and the global batch size is 512. The only difference is the supervision signal. Continued SFT uses teacher-forced cross-entropy. OPSD uses the KL objective with a frozen teacher initialized from the SFT checkpoint. Student outputs are sampled with temperature $\tau=0.8$. No auxiliary cross-entropy term is used.

All reported CER numbers use the same greedy decoding configuration unless otherwise noted. This setting is shared by CPT, SFT, OPSD, Direct Dialect SFT, and Continued SFT. Thus, differences reflect training rather than search. The sampling temperature $\tau=0.8$ is used only during OPSD training to construct on-policy prefixes. It does not change the evaluation decoder.

We will release evaluation scripts for the 13 evaluation sets. We will also release the hyperparameter settings, decoding settings, and random seeds used for the reported run.

\section{Results}
\label{sec:results}

\subsection{Main Results}

To evaluate the framework, Table~\ref{tab:summary_results} reports aggregate CER for open-source baselines and our staged checkpoints. Tables~\ref{tab:mandarin} and~\ref{tab:aux_dial} expand the 13 evaluation sets into Mandarin and dialect details, and Table~\ref{tab:dialect} covers the Internal Dialect sets. GLM/Fun appear in Table~\ref{tab:summary_results}.

\begin{table}[t]
\caption{Aggregate CER (\%) for open-source baselines and our staged checkpoints. Lower is better. The four average columns are defined in Table~\ref{tab:avg_definition}.}
\label{tab:summary_results}
\centering
\small
\setlength{\tabcolsep}{4pt}
\begin{tabular}{@{}lcccc@{}}
\toprule
\multirow{2}{*}{\textbf{Model}} & \textbf{Mandarin} & \textbf{Dialect} & \textbf{Internal} & \textbf{Overall} \\
 & \textbf{Avg.} & \textbf{Avg.} & \textbf{Avg.} & \textbf{Avg.} \\
\midrule
\multicolumn{5}{@{}l}{\textit{Open-source baselines}} \\
GLM-ASR & 5.28 & 37.21 & 41.55 & 31.49 \\
Fun-ASR & 4.59 & 18.43 & 27.10 & 19.89 \\
Qwen3-ASR & 3.46 & 15.37 & 21.01 & 15.57 \\
\midrule
\multicolumn{5}{@{}l}{\textit{Our staged checkpoints}} \\
CPT & 3.78 & 13.74 & 15.09 & 11.95 \\
SFT & 3.40 & 13.16 & 13.30 & 10.72 \\
OPSD & \textbf{3.27} & \textbf{12.79} & \textbf{12.42} & \textbf{10.12} \\
\bottomrule
\end{tabular}
\end{table}

\begin{table}[t]
\caption{CER (\%) on the eight Mandarin evaluation sets. Lower is better; bold marks the best staged model.}
\label{tab:mandarin}
\centering
\small
\setlength{\tabcolsep}{4pt}
\begin{tabular}{@{}lcccc@{}}
\toprule
\textbf{Evaluation set} & \textbf{Qwen3-ASR} & \textbf{CPT} & \textbf{SFT} & \textbf{OPSD} \\
\midrule
AISHELL-1 & 1.57 & 1.47 & 1.43 & \textbf{1.38} \\
AISHELL-2 & 2.79 & 2.62 & 2.58 & \textbf{2.52} \\
KeSpeech & 5.11 & 4.72 & 4.84 & \textbf{4.56} \\
SpeechIO-1 & 0.75 & \textbf{0.71} & 0.89 & 0.86 \\
SpeechIO-2 & 3.83 & 4.17 & 3.50 & \textbf{3.39} \\
SpeechIO-3 & 1.39 & 1.33 & 1.35 & \textbf{1.27} \\
Test\_Meeting & 6.74 & 8.50 & 7.22 & \textbf{6.85} \\
Test\_Net & 5.46 & 6.74 & 5.38 & \textbf{5.30} \\
\midrule
Mandarin Avg. & 3.46 & 3.78 & 3.40 & \textbf{3.27} \\
\bottomrule
\end{tabular}
\end{table}

\begin{table}[t]
\caption{CER (\%) on the five public dialect evaluation sets. Lower is better; bold marks the best staged model.}
\label{tab:aux_dial}
\centering
\small
\setlength{\tabcolsep}{4pt}
\begin{tabular}{@{}lccccc@{}}
\toprule
\textbf{Evaluation set} & \textbf{Dialect} & \textbf{Qwen3-ASR} & \textbf{CPT} & \textbf{SFT} & \textbf{OPSD} \\
\midrule
WenetSpeech-Yue Long & Cantonese & 9.99 & 9.40 & 9.01 & \textbf{8.80} \\
WenetSpeech-Yue Short & Cantonese & 6.93 & 6.09 & 5.60 & \textbf{5.31} \\
WenetSpeech-Chuan Easy & Sichuan & 12.38 & 12.91 & 12.30 & \textbf{11.86} \\
WenetSpeech-Chuan Hard & Sichuan & 21.79 & 23.24 & 22.27 & \textbf{21.74} \\
WenetSpeech-Wu & Wu & 25.74 & 17.06 & 16.60 & \textbf{16.26} \\
\midrule
Dialect Avg. & & 15.37 & 13.74 & 13.16 & \textbf{12.79} \\
\bottomrule
\end{tabular}
\end{table}\begin{table}[t]
\caption{CER (\%) on the 18 internal dialect evaluation sets. Lower is better; bold marks the best staged model.}
\label{tab:dialect}
\centering
\small
\setlength{\tabcolsep}{4pt}
\begin{tabular}{@{}lcccc@{}}
\toprule
\textbf{Dialect} & \textbf{Qwen3-ASR} & \textbf{CPT} & \textbf{SFT} & \textbf{OPSD} \\
\midrule
Anhui & 18.95 & 14.63 & 13.99 & \textbf{13.08} \\
Cantonese & 10.06 & 8.71 & 7.87 & \textbf{7.74} \\
Changsha & 14.79 & 11.98 & 10.83 & \textbf{10.23} \\
Chaoshan & 45.59 & 32.26 & 27.34 & \textbf{25.21} \\
Dongbei & 6.45 & 6.09 & 5.95 & \textbf{5.80} \\
Henan & 8.46 & 7.26 & 6.27 & \textbf{5.99} \\
Kejia & 60.47 & 37.70 & 32.01 & \textbf{28.60} \\
Minnan & 30.03 & 23.09 & 19.79 & \textbf{18.59} \\
Nanchang & 33.41 & 20.47 & 18.63 & \textbf{15.58} \\
Nanjing & 13.37 & 10.58 & 9.67 & \textbf{9.33} \\
Shanxi & 28.53 & 21.88 & 20.17 & \textbf{18.69} \\
Shaanxi & 9.68 & 6.93 & 6.69 & \textbf{6.28} \\
Shandong & 8.78 & 8.39 & 7.82 & \textbf{7.64} \\
Shanghai & 15.78 & 12.86 & \textbf{11.94} & 12.07 \\
Sichuan & 5.99 & 6.13 & \textbf{5.13} & 5.38 \\
Suzhou & 50.35 & 26.73 & 22.12 & \textbf{20.73} \\
Wuhan & 11.30 & 10.22 & 8.12 & \textbf{7.59} \\
Xuzhou & 6.12 & 5.66 & 5.10 & \textbf{5.04} \\
\midrule
Internal Avg. & 21.01 & 15.09 & 13.30 & \textbf{12.42} \\
\bottomrule
\end{tabular}
\end{table}

We read the results in the same order as the training story. Starting from the released Qwen3-ASR model, the full CPT--SFT--OPSD path lowers Mandarin Avg. CER from 3.46\% to 3.27\%, Dialect Avg. CER from 15.37\% to 12.79\%, and Internal Avg. CER from 21.01\% to 12.42\%.

\textbf{CPT.} CPT mainly improves the ASR foundation for later dialect training. Compared with Qwen3-ASR, it reduces Dialect Avg. CER from 15.37\% to 13.74\% and Internal Avg. CER from 21.01\% to 15.09\%. Mandarin Avg. CER rises from 3.46\% to 3.78\%, so CPT alone is not sufficient.

\textbf{SFT.} Dialect SFT further lowers Dialect Avg. CER from 13.74\% to 13.16\% and Internal Avg. CER from 15.09\% to 13.30\%. It also lowers Mandarin Avg. CER relative to CPT, but some Mandarin sets, such as KeSpeech and SpeechIO-1, remain worse than CPT. This shows that training with more dialect speech can change the Mandarin--dialect CER trade-off.

\textbf{OPSD.} OPSD improves all three evaluation groups after SFT. Mandarin Avg. CER improves from 3.40\% to 3.27\%, Dialect Avg. CER from 13.16\% to 12.79\%, Internal Avg. CER from 13.30\% to 12.42\%, and Overall Avg. CER from 10.72\% to 10.12\%. Relative to SFT, it lowers CER on all eight Mandarin sets and all five public dialect sets. Absolute gains include KeSpeech from $4.84\%$ to $4.56\%$, SpeechIO-2 from $3.50\%$ to $3.39\%$, Test\_Net from $5.38\%$ to $5.30\%$, and, on the internal suite, Nanchang from $18.63\%$ to $15.58\%$, Kejia from $32.01\%$ to $28.60\%$, and Chaoshan from $27.34\%$ to $25.21\%$. SpeechIO-1 is $0.86\%$ under OPSD, still above the CPT value of $0.71\%$. Section~\ref{sec:ablation} shows that these gains do not come from simply training one more epoch with cross-entropy.

A central question is whether lower dialect CER comes with higher Mandarin Avg. CER. We therefore keep the four averages (Mandarin Avg., Dialect Avg., Internal Avg., and Overall Avg.) separate.

Relative to SFT, OPSD lowers all four averages and all eight Mandarin sets in Table~\ref{tab:mandarin}. SpeechIO-1 remains worse than CPT, with CER of $0.86\%$ versus $0.71\%$, so OPSD is not best on every individual set. Its main value is the better Mandarin--dialect balance: unlike Continued SFT in Section~\ref{sec:ablation}, it does not raise Mandarin Avg. CER while refining dialect performance.

\subsection{Analysis of Dialect Performance}

Table~\ref{tab:dialect} shows a clear easy--hard split across the 18 internal dialects. Under OPSD, Dongbei, Sichuan, Xuzhou, and Henan stay below $6\%$ CER, while Shaanxi, Shandong, Wuhan, and Nanjing also remain relatively easy. In contrast, Kejia, Chaoshan, and Suzhou remain the hardest dialects, with CER of $28.60\%$, $25.21\%$, and $20.73\%$, followed by Minnan, Shanxi, and Nanchang. Cantonese reaches $7.74\%$ CER and is mid-tier rather than hard, despite being a southern variety.

These gaps align with several factors that often co-occur. First, distance from Mandarin: northern and Mandarin-continuum varieties such as Dongbei, Henan, Xuzhou, and Shaanxi share more phonetic and lexical overlap with the Mandarin-heavy pretraining and CPT mixture, so recognition starts from a stronger base. More distant Min, Hakka, and Wu varieties such as Minnan, Chaoshan, Kejia, and Suzhou diverge more in pronunciation and word choice, leaving a larger residual after adaptation. Second, public data coverage is uneven. Large transcribed resources such as WenetSpeech-Chuan and WenetSpeech-Yue directly support Sichuan and Cantonese-style speech, whereas Hakka and Chaoshan have less dedicated large-scale public coverage in our mixture. WenetSpeech-Wu helps Wu overall, but local cities such as Suzhou can remain difficult. Third, accent and speaking style still matter within a family: Shanghai is easier than Suzhou in our suite, and Sichuan is easier than many southern dialects, consistent with stronger resource support and milder mismatch to Mandarin ASR.

Across stages, CPT usually gives the largest single drop for the hardest dialects. For example, Kejia falls from $60.47\%$ to $37.70\%$, and Suzhou falls from $50.35\%$ to $26.73\%$. OPSD then adds further gains where early errors are more likely to accumulate: Nanchang from $18.63\%$ to $15.58\%$, Kejia from $32.01\%$ to $28.60\%$, and Chaoshan from $27.34\%$ to $25.21\%$. A few easier sets such as Shanghai and Sichuan are slightly better under SFT than OPSD, so the refinement is not uniformly best on every dialect. Overall, the pattern supports CPT for building dialect coverage and OPSD as a final refinement on difficult varieties.

\section{Ablation Studies}
\label{sec:ablation}

We use two ablations to check the main claims. The first asks whether CPT is useful before dialect SFT. The second asks whether OPSD is better than continuing cross-entropy training on the same data. Tables~\ref{tab:cpt_ablation} and~\ref{tab:refinement_ablation} report the two comparisons separately.

\begin{table}[t]
\caption{CER (\%) on the ablation of CPT before dialect SFT. Lower is better; bold marks the best.}
\label{tab:cpt_ablation}
\centering
\small
\setlength{\tabcolsep}{3pt}
\begin{tabular}{@{}lcccccc@{}}
\toprule
\multirow{2}{*}{\textbf{Model}} & \multirow{2}{*}{\textbf{CPT}} & \multirow{2}{*}{\textbf{SFT}} & \textbf{Mandarin} & \textbf{Dialect} & \textbf{Internal} & \textbf{Overall} \\
 & & & \textbf{Avg.} & \textbf{Avg.} & \textbf{Avg.} & \textbf{Avg.} \\
\midrule
Qwen3-ASR & -- & -- & 3.46 & 15.37 & 21.01 & 15.57 \\
w/o CPT & -- & \checkmark & 5.16 & 14.70 & 14.20 & 11.95 \\
w/ CPT & \checkmark & \checkmark & \textbf{3.40} & \textbf{13.16} & \textbf{13.30} & \textbf{10.72} \\
\bottomrule
\end{tabular}
\end{table}

\begin{table}[t]
\caption{CER (\%) on the final refinement step. Continued SFT and OPSD start from the same SFT checkpoint. Lower is better; bold marks the best.}
\label{tab:refinement_ablation}
\centering
\small
\setlength{\tabcolsep}{4pt}
\begin{tabular}{@{}lcccc@{}}
\toprule
\multirow{2}{*}{\textbf{Model}} & \textbf{Mandarin} & \textbf{Dialect} & \textbf{Internal} & \textbf{Overall} \\
 & \textbf{Avg.} & \textbf{Avg.} & \textbf{Avg.} & \textbf{Avg.} \\
\midrule
SFT checkpoint & 3.40 & 13.16 & 13.30 & 10.72 \\
\hspace{1em}+ Continued SFT & 4.43 & 12.89 & 12.95 & 10.74 \\
\hspace{1em}+ OPSD & \textbf{3.27} & \textbf{12.79} & \textbf{12.42} & \textbf{10.12} \\
\bottomrule
\end{tabular}
\end{table}

\subsection{Effect of CPT}

Table~\ref{tab:cpt_ablation} compares the Qwen3-ASR, dialect SFT without CPT, and dialect SFT after CPT. Direct dialect SFT lowers Dialect Avg. CER from 15.37\% to 14.70\% and Internal Avg. CER from 21.01\% to 14.20\%, but substantially raises Mandarin Avg. CER from 3.46\% to 5.16\%. Overall Avg. CER falls from 15.57\% to 11.95\% under the macro-average, yet this is not a balanced gain because Mandarin Avg. CER increases substantially. Adding CPT before SFT yields 3.40\% Mandarin Avg. CER, 13.16\% Dialect Avg. CER, 13.30\% Internal Avg. CER, and 10.72\% Overall Avg. CER. Thus CPT on Mandarin-dialect data before dialect SFT is needed for a stable Mandarin--dialect trade-off and provides the SFT checkpoint used by OPSD.

\subsection{Effect of OPSD}

Table~\ref{tab:refinement_ablation} starts from the same SFT checkpoint. Continued SFT slightly lowers Dialect Avg. CER from 13.16\% to 12.89\% and Internal Avg. CER from 13.30\% to 12.95\%, but raises Mandarin Avg. CER from 3.40\% to 4.43\%; Overall Avg. CER rises from 10.72\% to 10.74\%. OPSD lowers all three group averages and Overall Avg. CER to 10.12\%. Both runs use the same $\mathcal{D}_{\text{ref}}$ and schedule (Section~\ref{sec:impl}), so the difference is the objective rather than extra data or training budget. Relative to Continued SFT, OPSD lowers Mandarin Avg. CER from 4.43\% to 3.27\%, Dialect Avg. CER from 12.89\% to 12.79\%, Internal Avg. CER from 12.95\% to 12.42\%, and Overall Avg. CER from 10.74\% to 10.12\%.

\section{Conclusion}

%
%
\bibliographystyle{splncs04}
\bibliography{bib}

\end{document}